\documentclass[aps,pra,twocolumn,10pt,nofootinbib]{revtex4-2}
\usepackage{bm,dcolumn,amsmath,graphicx,amsfonts,amssymb,physics}

\usepackage{hyperref,xcolor}
\definecolor{citecol}{rgb}{0.,0.,0.9}
\hypersetup{colorlinks,linkcolor={citecol},citecolor={citecol},urlcolor={citecol}}

\renewcommand{\v}[1]{\ensuremath{\boldsymbol{#1}}} 

\AtBeginDocument{\DeclareMathSymbol{:}{\mathord}{operators}{"3A}}

\begin{document}

\title{Breit corrections to moderately charged ions in all-orders calculations}

\date{19 February 2026} 

\author{Andoni Skoufris}

\author{Benjamin M. Roberts}\email[]{b.roberts@uq.edu.au}

\affiliation{School of Mathematics and Physics, The University of Queensland, Brisbane QLD 4072, Australia}

\begin{abstract}
\noindent
The atomic properties of heavy, moderately-charged ions are important for a wide variety of applications, 
including precision tests of fundamental physics and for the study and development of atomic and nuclear clocks.
In these systems it is known that relativistic effects, such as the Breit interaction and radiative quantum electrodynamics corrections, are important for an accurate understanding of atomic properties.
It is also known that inclusion of correlations alongside the Breit effect is crucial.
In this work we include the Breit interaction into all-orders calculations of energy levels and fine structure intervals of ions in the Cs and Fr isoelectronic sequences.
This requires modifying the electron Green's function to account for Breit within the all-orders correlation potential method, which sums dominating series of perturbation diagrams exactly using a Feynman diagram technique.
We find that Breit corrections to the energies of moderately ionized ions along these sequences are very large, particularly for the $f$ states.
We also observe a significant deviation from experiment for these levels. Incorporating Breit into the all-orders correlation potential provides a significant additional contribution beyond including Breit at the second-order level alone.
While this does not resolve the disagreement in the energy levels, it does substantially improve the fine-structure intervals beyond what is achieved by including Breit only at second order.
Furthermore, we include the frequency-dependent Breit interaction into the Dirac-Fock procedure, and find that this does not significantly modify the energy levels at this order of approximation.
\end{abstract}

\maketitle

\section{Introduction}
An accurate theoretical understanding of atomic systems has far-reaching applications in many areas of physics.
For example, high-precision calculations of atomic parity violating processes inside atoms provides one of the most stringent tests of new physics beyond the Standard Model, and one of the highest precision tests of the electroweak sector of the Standard Model (SM) in general, see, e.g., Refs.~\cite{Ginges:2003qt, Roberts:2014bka, Safronova:2017xyt}.
In this context, it is known that the common starting point for atomic structure calculations, the Dirac-Fock approximation, is not sufficient on its own for high accuracy calculations, and that correlation and relativistic effects must be taken into account. 
The Breit interaction gives the leading order relativistic correction to the Coulomb interaction and is one of the most important relativistic effects inside a heavy atom such as Cs.
For example, combining experiment~\cite{Wood:1997zq} with calculations~\cite{Dzuba:1989bor,Blundell:1991.43.3407} of atomic atomic parity violation in Cs indicated a 2.5\,$\sigma$ deviation from the Standard Model.
Derevianko~\cite{Derevianko:2000dt,Derevianko:2001gtm} and others \cite{Sushkov:2000ja, Dzuba:2001.63.044103, Kozlov:2001bn} demonstrated that the Breit effect (along with radiative QED corrections~\cite{Johnson:2001nk, Kuchiev:2002fg, Milstein:2002ai, Flambaum:2005ni}) resolved this disagreement. 

Current calculations of atomic parity violation indicate reasonable agreement with the SM \cite{Dzuba:2002kx, Porsev:2009pr,*Porsev:2010de, Dzuba:2012kx}.
Calculations typically only include the one-body part of the Breit corrections at the level of second order in the Coulomb interaction. 
While this level of inclusion clearly leads to very satisfactory theoretical predictions in Cs, there is currently ongoing work towards studying parity violation in Fr, Ra$^+$ and heavier ions~\cite{NunezPortela:2014.114.173, Gwinner:2022ijo, Roberts:2013fxq}, and there have also recently been several important experimental breakthroughs towards the construction of the first nuclear clock using the nucleus of a thorium atom 
\cite{Tiedau:2024obk, Zhang:2024ngu, Elwell:2024qyh}. 
This further motivates precision atomic-structure and symmetry-violation studies in heavy and highly charged systems, including proposals to exploit highly charged ions as sensitive probes of fundamental physics~\cite{Kozlov:2018mbp}. 
In parallel, complementary efforts on atomic clocks based on heavy ions~\cite{Versolato:2011.83.043829, Holliman:2022.128.033202, Cserveny:2025bmn} highlight the broader potential of these platforms for precision metrology and tests of fundamental symmetries.

In this work we will demonstrate that in such ions the deviation from experiment for the theoretical energies of $f$ states are unusually large.
We will show that the Breit corrections to these levels are very large, being the right order of magnitude to be the cause of this issue.
This problem is particularly relevant for the nuclear clock project, as one way to investigate the long-term viability of thorium for a nuclear clock is through highly precise studies of its hyperfine structure.
One accessible way to do this is through studies of the Th$^{3+}$ ion, whose ground state is $5f_{5/2}$, and so being able to accurately predict the $f$ states in heavy, moderately charged ions would be important in such studies.

Due to the importance of treating Breit and second order correlations concurrently, including Breit and all-orders correlations on the same footing may also lead to important corrections in cases where Breit and all-orders corrections are large, such as heavy ions.
As such, in this work we include the one-body part of the Breit interaction into the Green's function in the all-orders correlation method developed in Refs.~\cite{Dzuba:1988.131.461, Dzuba:1989.140.493}, thereby including the Breit interaction into an important series of non-perturbative correlation effects to all orders (we stress that we mean all orders in the residual Coulomb interaction; Breit should be treated to first order only).
We also show that inclusion of Breit into the all-orders correlation potential method greatly improves the fine structure intervals, leading to near-perfect agreement with experiment. However, the large disagreement between theory and experiment for the energy levels still remains.
We also include higher order relativistic corrections into the Dirac-Fock equation, in the form of the frequency-dependent Breit interaction, however, we find that the resulting corrections are very small.

\section{Theory}

The starting point for modeling single-valence atomic systems is the Dirac-Fock equation,
\begin{align}
    \left(h_0 + V_\mathrm{DF}\right)\psi_a&=\varepsilon_a\psi_a,
\end{align}
where $h_0$ is the single-particle Dirac Hamiltonian including the nuclear potential and $V_\mathrm{DF}$ is the usual frozen-core Dirac-Fock potential due to the $N-1$ core electrons.
Correlations between the core and valence electrons can be taken into account by adding a non-local, energy-dependent operator $\Sigma(\varepsilon)$, known as the correlation potential, to the Dirac-Fock equation for the valence electrons, 
\begin{align}
    \left[h_0 + V_\mathrm{DF} + \Sigma(\varepsilon)\right]\psi^\mathrm{(BO)}_a&=\varepsilon^\mathrm{(BO)}_a\psi_a^\mathrm{(BO)},
\end{align}
whose expectation value is the correlation correction to the energy from treating the residual electron-electron interactions perturbatively. We may calculate $\Sigma$ to lowest (second) order or to all-orders, as discussed below. The solutions to this equation are known as Brueckner orbitals and the self-consistency of this equation includes an additional correction, known as the chaining of the self-energy operator, to all orders \cite{Dzuba:1984rva,Dzuba:1985.18.597}.

\begin{figure}
    \centering
    \includegraphics[width=0.33\textwidth]{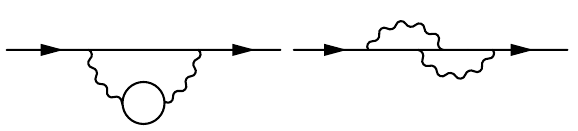}
    \caption{Feynman diagrams representing the direct (left) and exchange (right) contributions to the second order correlation correction to the energy of a single valence state.}
    \label{fig:Sigma2Feynman}
\end{figure}

The conventional way to construct the Dirac-Fock and correlation potentials is to assume that the electrons interact simply through the instantaneous Coulomb interaction. 
However, for high accuracy calculations in highly relativistic systems, relativistic and QED corrections to the electron-electron interactions need to be introduced.
The Coulomb interaction arises as the effective interaction from the exchange of a single virtual photon in $e^-e^-\to e^-e^-$ scattering at second order in QED, when calculated to lowest order in powers of $v/c$.
In the Coulomb gauge, this amounts to only retaining the temporal components of the photon propagator and neglecting the exchange of transverse photons between the electrons.
When the transverse components of the propagator are kept, but only to order $(v/c)^2$, the resulting effective operator,
\begin{align}
    B(\bm{r}_1,\bm{r}_2)&=-\frac{e^2}{2r_{12}}\left[\v{\alpha}_1\!\cdot\!\v{\alpha}_2 + \frac{(\v{\alpha}_1\!\cdot\!\v{r}_{12})(\v{\alpha}_2\!\cdot\!\v{r}_{12})}{r_{12}^2}\right], \label{eq:breit-interaction}
\end{align}
is known as the Breit interaction, where $\v{\alpha}_i$ is a Dirac matrix which acts on the $i$th electron and $\v{r}_{12}$ is the separation vector between the two electrons. 
In the Feynman gauge, the interpretation of the Breit interaction is instead as the lowest order relativistic correction to the Coulomb interaction when treated in powers of $v/c$ \cite{Greiner:2009.10.1007/978-3-540-87561-1}.

For high precision calculations of relativistic systems it is also important to include QED radiative effects due to vacuum polarization and electron self-energy, which we include with the QED radiative potential method \cite{Flambaum:2005ni}. In this method, an effective local potential is added to the Hamiltonian, which allows us to include QED radiative corrections into atomic wave functions.

\begin{figure}[t!]
\centering
    \includegraphics[width=0.47\textwidth]{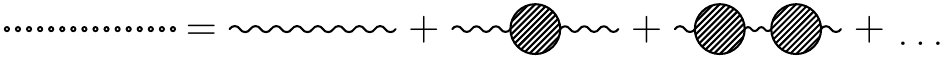}
    \includegraphics[width=0.4\textwidth]{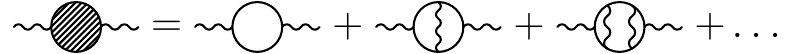}
    \caption{\small All-orders screening of the Coulomb operator (top) and hole-particle interaction (bottom).}
    \label{fig:screening}
\end{figure}

\subsection{Breit into correlations: second-order}

To more accurately treat relativistic systems, the Breit interaction should be included into atomic structure calculations, which can be done in two main ways.
The first way is to solve for the core and valence wave functions with the usual DF potential and then simply treat the Breit interaction as a perturbation, in which case the second-quantized Hamiltonian reads~\cite{Derevianko:2001gtm}
\begin{align}
\begin{split}
    H_C&=\sum_{i}\varepsilon_i:a^\dagger_i a_i: + \sum_{ij}b_{ij}:a^\dagger_ia_j: \\
    &\hspace{6em}+ \frac{1}{2}\sum_{ijkl}(g_{ijkl} + b_{ijkl}):a^\dagger_ia^\dagger_ja_la_k:, \label{eq:coulomb-ham}
\end{split}
\end{align}
where $a_i$ ($a_i^\dagger$) removes (adds) an electron in the state $\psi_i$ to the Slater determinant wave function, $g_{ijkl}$ is a two-body Coulomb matrix element, $b_{ijkl}$ is a two-body Breit matrix element, and $g_{ij}=\sum_a^{\mathrm{core}}(g_{iaja} - g_{iaaj})$ is an effective one-body Coulomb matrix element, and likewise for $b_{ij}$. 
The second way is to include the Breit interaction at the level of the DF equations by adding the one-body effective Breit operator $(V_{\mathrm{Br}})_{ij}=b_{ij}$ to the DF potential. Doing so modifies the core and excited wave functions at the level of the DF equations, and results in the Hamiltonian,
\begin{align}
    H_{C+B}&=\sum_{i}\bar{\varepsilon}_i:\bar{a}^\dagger_i \bar{a}_i: + \frac{1}{2}\sum_{ijkl}(\bar{g}_{ijkl} + \bar{b}_{ijkl}):\bar{a}^\dagger_i\bar{a}^\dagger_j\bar{a}_l\bar{a}_k: \label{eq:breit-coulomb-ham}
\end{align}
where we have adopted the notation of Ref.~\cite{Derevianko:2001gtm} and denoted Breit-Dirac-Fock (BDF) orbitals, wave functions, energies, etc.\ with overbars and the usual DF counterparts with no bars. In the first approach, (\ref{eq:coulomb-ham}), there will be corrections to the diagonal part of the Hamiltonian from both the one-body and two-body parts of the Breit interaction, which can be treated order-by-order in perturbation theory. However, in the second approach, (\ref{eq:breit-coulomb-ham}), the iteration of the DF equations with the inclusion of $V_{\mathrm{Br}}$ captures an infinite series of diagrams that sum the one-body part of the Breit interaction to all orders, and hence the one-body Breit terms do not appear in the Hamiltonian in Eq.\ (\ref{eq:breit-coulomb-ham}). This means that by including Breit at the level of the DF procedure, we capture a dominant series of diagrams that ensure that first order corrections due to the Breit interaction identically vanish, which also makes the perturbation procedure much more numerically simple. 

Importantly, we note that while this includes the Breit interaction into diagrams that are non-linear in Breit, the change in the effect of the Breit interaction at each order does not stem from these terms.
Any terms that are non-linear in the Breit interaction are strictly nonphysical, and their effect can be understood by numerically calculating $\partial\varepsilon/\partial\lambda$, where $\varepsilon$ is the energy shift caused by Breit and $\lambda$ is a scaling parameter for the Breit operator (see also Ref.~\cite{Derevianko:2001gtm} for discussion and an alternative treatment that is strictly linear in Breit). We find that the resulting corrections are linear in $\lambda$, and so the nonlinear terms do not contribute. 

It has been demonstrated in Refs.\ \cite{Derevianko:2001gtm,Lindroth:1989.22.2447,Kozlov:2000.arXiv:physics/0004076} that the corrections to energies and matrix elements that come from including Breit into the DF potential can differ in order of magnitude or sign than corrections that come simply from treating it as a perturbation at the same (second) order of perturbation theory. This is due to the relaxation effect that is accounted for when including $V_{\mathrm{Br}}$ into the DF procedure, as this modifies the core and excited wave functions (and therefore the DF potential itself) that are then used in many-body perturbation theory (MBPT) expressions. Therefore, it is important that we include the Breit interaction at the level of the DF potential when calculating correlation corrections, as this leads to the dominant contribution from the Breit interaction in the calculation of correlations. In this work, we follow Ref.~\cite{Derevianko:2001gtm} and refer to corrections that come from the inclusion of Breit into the DF potential as one-body Breit corrections.

At the level of second order correlations, one-body Breit corrections are accounted for by adding $V_{\mathrm{Br}}$ into the DF potential when forming the basis that is used in second order MBPT expressions. Although it is very small, we further include the two-body part of the Breit interaction using the regular MBPT expressions for second order correlations but taking two-body Coulomb matrix elements to Coulomb + Breit matrix elements, $g_{ijkl}\to g_{ijkl}+b_{ijkl}$, as outlined in Ref.~\cite{Johnson:1988.37.2764}. 

\subsection{Breit into all-orders correlations}

Because of the large effect that Breit plays at second order in correlations, including it into higher order correlations may lead to important additional corrections. Although the correlation potential can be calculated in an order-by-order expansion in the Coulomb interaction, for accurate calculations all-orders methods are required. As such, in this work we have included the Breit interaction into the Feynman diagram method of perturbation theory, which allows us to include the one-body part of Breit into an important series of many-body diagrams which are summed exactly to all orders in the Coulomb interaction. At second order, the Goldstone and Feynman methods are physically equivalent, with the direct and exchange diagrams in the Feynman method being shown in Figure \ref{fig:Sigma2Feynman}. For the direct diagram, this translates into the expression,
\begin{align}
\begin{split}
    \Sigma^{(2)}_{\mathrm{d}}(\bm{r}_1,\bm{r}_2;\varepsilon)&=\int d^3r_i\,d^3r_j\int\frac{d\omega}{2\pi}\,G_{12}(\varepsilon+\omega)\\
    &\hspace{8.8em}\times Q_{1i}\Pi_{ij}(\omega)Q_{j2} , \label{eq:F2-direct}
\end{split}
\end{align}
where $G_{ij}(\varepsilon)\equiv G(\v{r}_i,\v{r}_j;\varepsilon)$ is the time-ordered Feynman Green's function for the Dirac-Fock operator, satisfying,
\begin{align}
    (\varepsilon - h_0 - V^{\mathrm{DF}})G(\bm{r}_1,\bm{r}_2;\varepsilon)=\delta^3(\bm{r}_1-\bm{r}_2),
\end{align}
$Q_{ij}\equiv |\bm{r}_{i} - \bm{r}_j|^{-1}$ is the Coulomb operator,
\begin{align}
\Pi(\bm{r}_1,\bm{r}_2;\omega)=\int\frac{d\varepsilon}{2\pi}\mathrm{Tr}[G(\bm{r}_1,\bm{r}_2;\varepsilon)G(\bm{r}_2,\bm{r}_1;\varepsilon+\omega)] \label{eq:polarisation-operator}
\end{align}
is the (core) polarization operator, and the trace is over spinor indices. 

\begin{figure}[!t]
    \centering
    \includegraphics[width=0.33\textwidth]{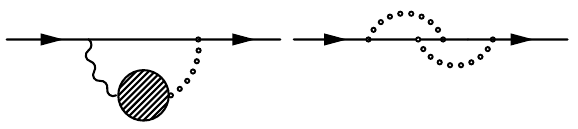}
    \caption{Feynman diagrams for the direct (left) and exchange (right) all-orders correlation corrections.}
    \label{fig:all-orders-diagram}
\end{figure}

In the all-orders correlation potential method, we include an infinite series of screening diagrams by exactly calculating the series of diagrams shown in Fig.\ \ref{fig:screening}, as outlined in Refs.\ \cite{Dzuba:1988.131.461,Dzuba:1989.140.493,Roberts:2022lda}. The first and most important of these corresponds to the screening of the Coulomb interaction by the atomic core, while the second is known as the hole-particle interaction and corresponds physically to the deviation of the excited electron in the hole-particle loop from the $V^{N-1}$ potential. 
The all-orders correlation potential is obtained by substituting these series of diagrams into the second-order Feynman diagrams, as shown in Fig.~\ref{fig:all-orders-diagram}, thereby capturing a dominating series of correlation corrections exactly. Due to the numerically expensive double frequency integral in the exchange term and its relative smallness to the direct term we only calculate the direct Feynman diagram, and instead calculate the all-orders exchange diagram using the Goldstone method with screening factors, as outlined in Ref.~\cite{Dzuba:1988.131.461}.

The Green's function for local potentials (e.g.\ DF excluding exchange), $G_0$ can be calculated exactly (without the use of a basis) using standard Green's function techniques. 
The non-local exchange potential can then be taken into account non-perturbatively by solving the Dyson equation,
\begin{align}
    G=(1-G_0V_\mathrm{x})^{-1}G_0,
\end{align}
where $V_{\rm DF} = V_{\mathrm{l}} + V_{\mathrm{x}}$ includes the local direct potential and non-local exchange term. In order to include the one-body part of the Breit interaction into correlations, we need to replace the DF Green's function in Eq.\ (\ref{eq:F2-direct}) and the exchange diagram with the Breit-Dirac-Fock (BDF) Green's function, $\bar{G}$, which satisfies
\begin{align}
    (\varepsilon - h_0 - V_{\mathrm{DF}} - V_{\mathrm{Br}})\bar{G}(\bm{r}_1,\bm{r}_2;\varepsilon)=\delta^3(\bm{r}_1-\bm{r}_2)
\end{align}
Because the effective one-body Breit operator $V_{\mathrm{Br}}$ is just an addition to the DF potential we can include it into the Green's function using the Dyson equation in exactly the same manner as we do for the exchange potential:
\begin{align}
    \bar{G}&=[1 - G_0(V_\mathrm{x} + V_\mathrm{Br})]^{-1}G_0.
\end{align}
Constructing the Feynman diagrams with $\bar{G}$ rather than $G$ thereby includes the one-body part of the Breit interaction into the correlation potential.

Including Breit into the Green's function through the Dyson equation involves constructing a coordinate representation of $V_\mathrm{Br}$ which can be done either directly from Eq.\ (\ref{eq:breit-interaction}) or by using a complete set of wave functions,
\begin{align}
    V_\mathrm{Br}(\v{r},\v{r}')&=\sum_i\langle\bm{r}|V_{\mathrm{Br}}|i\rangle\langle i|\bm{r}'\rangle=\sum_i[V_\mathrm{Br}\psi_i](\v{r})\psi^\dagger_i(\v{r}').
\end{align}
In practice, we perform this summation using 90 hydrogenic basis states over the partial waves $l=0$ to $l=6$ in a cavity of radius 90\,a.u. This also allows us to easily account for the spinor structure of $V_\mathrm{Br}$, as discussed below.

In this work we use the Dirac representation and assume separable wave functions of the form,
\begin{align}
    \psi_{n\kappa m}(\bm{r}) = \frac{1}{r}
    \begin{pmatrix}
        f_{n\kappa}(r) \Omega_{\kappa m}(\bm{n}) \\
        i g_{n\kappa}(r) \Omega_{-\kappa,m}(\bm{n})
    \end{pmatrix},
\end{align}
where $f$ and $g$ are respectively the large and small parts of the wave function, and $\Omega$ is a two-component spinor. Practically, the angular integrations can be carried out analytically and we only have to numerically calculate the radial integrals. Similarly, the Green's function and exchange (and Breit) potential can be represented by radial spinor matrices, whose components may be denoted by
\begin{align}
    G(r_i,r_j)&\equiv\begin{pmatrix} G_{ff}(r_i,r_j) &  G_{fg}(r_i,r_j)  \\ G_{fg}(r_i,r_j) &  G_{gg}(r_i,r_j) \end{pmatrix},
\end{align} 
and unlike the Coulomb interaction, the Breit interaction mixes the relativistic and non-relativistic spinor components of wave functions. Symbolically, the exchange potential has the structure
\begin{align}
    V_\mathrm{x}(r,r')&\sim\sum_{b}\begin{pmatrix} f_b(r)f_b(r') & f_b(r)g_b(r') \\ g_b(r)f_b(r') & g_b(r)g_b(r') \end{pmatrix},
\end{align}
while the one-body effective Breit operator is of the form
\begin{align}
    V_\mathrm{Br}(r,r')&\sim\sum_{b}\begin{pmatrix} g_b(r)f_b(r') & g_b(r)g_b(r') \\ \pm f_b(r)f_b(r') & f_b(r)g_b(r') \end{pmatrix},
\end{align}
We can see that for the exchange potential, the $ff$ part of the exchange potential is the largest, while for the one-body Breit operator, the most important contributions are in the off-diagonal ($fg$ and $gf$) components. It's often a reasonable approximation to only include the non-relativistic ($ff$) part of the Green's function, however, this would miss the largest contributions from the one-body Breit operator. Therefore, we have used the full Green's function when solving the Dyson equation with the Breit interaction. Furthermore, because the polarization operator contains a trace over the spinor indices of the product of two Green's functions, we include all the components of the Green's functions in this trace.

It should be noted that with the use of BDF orbitals, the two-body Breit corrections will first be non-vanishing at second order in the residual electron-electron interactions [see Eq.\ (\ref{eq:breit-coulomb-ham})]. Because the two-body Breit corrections are known to be an order of magnitude or more smaller than the one-body corrections, we do not include them into our construction of the all-orders correlation potential, however, we include them at the second order level using the Goldstone method (see Refs.\ \cite{Blundell:1987hmf,Mann:1971.4.41}).

\section{Results and Discussion}
\begin{table*}
\caption{Comparison of theoretical ionization energies of Fr to experimental energies (cm$^{-1}$), with the breakdown of contributions beyond the Dirac-Fock (DF) level from the all-orders correlations $\Sigma^{(\infty)}$, Breit corrections at the second ($\Sigma^{(2)}$) and all-orders ($\Sigma^{(\infty)}$) levels, and QED corrections. The column $\delta B(\Sigma^{(2)})$ shows the Breit correction from including the one-body and two-body parts of the Breit interaction into $\Sigma^{(2)}$, while $\delta B(\Sigma^{(\infty)})$ shows the additional correction from including the Breit interaction into the Green's function in $\Sigma^{(\infty)}$. The total theoretical energy is the sum of each column.}
\label{tab:francium-energies}
\begin{ruledtabular}
\begin{tabular}{@{}lD{.}{.}{5.1}D{.}{.}{4.1}D{.}{.}{3.1}D{.}{.}{2.1}D{.}{.}{3.1}D{.}{.}{5.1}D{.}{.}{5.1}D{.}{.}{3.1}D{.}{.}{2.2}@{}}
Fr & \multicolumn{1}{c}{\rm{DF}} & \multicolumn{1}{c}{$\delta\Sigma^{(\infty)}$} & \multicolumn{1}{c}{$\delta B(\Sigma^{(2)})$} & \multicolumn{1}{c}{$\delta B(\Sigma^{(\infty)})$} & \multicolumn{1}{c}{\rm{QED}} & \multicolumn{1}{c}{\rm{Total}} & \multicolumn{1}{c}{\rm{Expt}.\ \cite{NIST_ASD}} & \multicolumn{1}{c}{$\Delta$} & \multicolumn{1}{c}{$\Delta\ (\%)$} \\ 
\hline
$7s_{1/2}$ & 28767.1 & 4076.3 & 5.7   & -5.4 & -45.3 & 32798.3 & 32848.9 & -50.6 & -0.15 \\
$7p_{1/2}$ & 18855.2 & 1769.9 & -14.4 & -1.2 & 0.9   & 20610.3 & 20611.5 & -1.1  & -0.01 \\
$7p_{3/2}$ & 17655.3 & 1255.2 & -0.5  & -1.6 & -0.4  & 18908.0 & 18924.9 & -16.9 & -0.09 \\
$6d_{3/2}$ & 13825.5 & 2796.2 & 33.9  & -4.1 & 11.5  & 16663.1 & 16619.0 & 44.1  & 0.27  \\
$6d_{5/2}$ & 13924.5 & 2515.8 & 37.0  & -4.1 & 8.9   & 16482.1 & 16419.2 & 62.9  & 0.38  \\
$8s_{1/2}$ & 12281.5 & 813.9  & -0.1  & -0.9 & -9.8  & 13084.6 & 13108.9 & -24.3 & -0.18 \\
$8p_{1/2}$ & 9240.3  & 494.3  & -4.8  & -0.4 & 0.3   & 9729.7  & 9735.9  & -6.3  & -0.06 \\
$8p_{3/2}$ & 8811.2  & 375.6  & -0.5  & -0.5 & -0.1  & 9185.8  & 9190.6  & -4.8  & -0.05 \\
\end{tabular}
\end{ruledtabular}
\end{table*}

\begin{table*}
\setlength{\tabcolsep}{6pt}
\caption{Comparison of theoretical ionization energies of La$^{2+}$ to experimental energies (cm$^{-1}$).}
\label{tab:lanthanum-energies}
\begin{ruledtabular}
\begin{tabular}{@{}lD{.}{.}{7.0}D{.}{.}{5.1}D{.}{.}{4.1}D{.}{.}{3.1}D{.}{.}{4.1}D{.}{.}{6.0}D{.}{.}{6.0}D{.}{.}{4.0}D{.}{.}{1.2}@{}}
           La$^{2+}$ & \multicolumn{1}{c}{\rm{DF}} & \multicolumn{1}{c}{$\delta\Sigma^{(\infty)}$} & \multicolumn{1}{c}{$\delta B(\Sigma^{(2)})$} & \multicolumn{1}{c}{$\delta B(\Sigma^{(\infty)})$} & \multicolumn{1}{c}{\rm{QED}} & \multicolumn{1}{c}{\rm{Total}} & \multicolumn{1}{c}{\rm{Expt}.\ \cite{NIST_ASD}} & \multicolumn{1}{c}{$\Delta$} & \multicolumn{1}{c}{$\Delta\ (\%)$} \\ \hline
$5d_{3/2}$ & 144973  & 10417.1                 & 62.6 & -2.1       & 35.6     & 155487     & 154675                      & 812    & 0.52         \\
$5d_{5/2}$ & 143712  & 10022.0                 & 110.2 & -3.9      & 29.4     & 153870     & 153072                      & 798    & 0.52         \\
$4f_{5/2}$ & 114362  & 33114.6                 & 748.1 & -29.0      & 119.1    & 148315     & 147480                      & 835    & 0.57         \\
$4f_{7/2}$ & 113259  & 32599.3                 & 889.6 & -40.2      & 109.5    & 146817     & 145980                      & 837    & 0.57         \\
$6s_{1/2}$ & 134360  & 6849.9                  & -17.3 & -5.2      & -68.1    & 141119     & 141084                      & 35     & 0.02         \\
$6p_{1/2}$ & 108013  & 4774.9                  & -55.4 & -1.9      & 6.0      & 112737     & 112660                      & 77     & 0.07         \\
$6p_{3/2}$ & 105315  & 4330.3                  & -19.0 & -2.8      & 0.4      & 109624     & 109564                      & 60     & 0.05         \\
$7s_{1/2}$ & 70190   & 2119.6                  & -8.3 & -1.6       & -23.3    & 72276      & 72328                       & -52    & -0.07        \\
$6d_{3/2}$ & 69863   & 2315.8                  & 8.1 & -1.7        & 6.4      & 72192      & 72294                       & -102   & -0.14        \\
$6d_{5/2}$ & 69464   & 2277.9                  & 19.2 & -2.1       & 5.2      & 71765      & 71861                       & -96    & -0.13        \\
\end{tabular}
\end{ruledtabular}
\end{table*}

\subsection{Energies}
Table \ref{tab:francium-energies} shows our theoretical energies for the lowest-lying valence states of francium. We can see that there is very good agreement with experiment for all levels (excepting the $6d$ levels). For these, the disagreement is due to these states' strong overlap with the core, meaning that valence-core interactions are particularly important. The energy levels can be improved with the inclusion of the so-called ladder diagrams \cite{Dzuba:2008.78.042502}, which account for these interactions. It is important to note that for practically all states, the most important Breit correction comes at the $\Sigma^{(2)}$ level [the $\delta B(\Sigma^{(2)})$ column], and the additional correction that comes from including Breit into $\Sigma^{(\infty)}$ [shown in the $\delta B(\Sigma^{(\infty)})$ column] leads in most cases to only a small correction. For some levels, such as $7s_{1/2}$, the additional correction from including Breit into $\Sigma^{(\infty)}$ is important, almost exactly canceling the correction at the second-order level. However, in these cases, the deviation from experiment is not significantly affected by this additional correction.

\begin{table*}[t!]
\setlength{\tabcolsep}{6pt}
\caption{Cumulative Breit and total QED corrections to La$^{2+}$ energy levels at each order of perturbation theory (cm$^{-1}$). The column labeled $\langle B\rangle$ shows the first order Breit correction, DF gives the correction to the DF energy when $V_{\mathrm{Br}}$ is included into the DF procedure, and the $\Sigma^{(2)}$ and $\Sigma^{(\infty)}$ columns are the corrections to the $\Sigma^{(2)}$ and $\Sigma^{(\infty)}$ energies from including just the one-body part of Breit into $\Sigma^{(2)}$ and $\Sigma^{(\infty)}$, respectively. Lastly, $B^{(2)}$ gives the correction to the corresponding energy due to the inclusion of the two-body part of the Breit interaction into $\Sigma^{(2)}$, calculated in the Goldstone method.}
\label{tab:lanthanum-breit-qed-corrections}
\begin{ruledtabular}\begin{tabular}{@{}lD{.}{.}{4.1}D{.}{.}{3.1}D{.}{.}{3.1}D{.}{.}{3.1}D{.}{.}{3.1}D{.}{.}{2.1}@{}}
& \multicolumn{5}{c}{Breit} & \mathrm{QED} \\ \cline{2-6}
& \multicolumn{1}{c}{$\langle B\rangle$} & \multicolumn{1}{c}{DF} & \multicolumn{1}{c}{$\Sigma^{(2)}$} & \multicolumn{1}{c}{$\Sigma^{(2)}+B^{(2)}$} & \multicolumn{1}{c}{$\Sigma^{(\infty)} + B^{(2)}$} & \\ \hline
$5d_{3/2}$ & -196.9                             & 48.5    & 57.3         & 62.6                 & 60.6       & 35.6            \\
$5d_{5/2}$ & -147.6                             & 88.0    & 105.1        & 110.2                & 106.3         & 29.4            \\
$4f_{5/2}$ & -406.5                             & 619.6   & 727.6        & 748.1                & 719.1        & 119.1           \\
$4f_{7/2}$ & -275.7                             & 722.9   & 868.5        & 889.6                & 849.4   & 109.5           \\
$6s_{1/2}$ & -116.3                             & -30.2   & -20.6        & -17.3                & -22.5     & -68.1           \\
$6p_{1/2}$ & -105.5                             & -58.5   & -57.9        & -55.4                & -57.3    & 6.0             \\
$6p_{3/2}$ & -75.4                              & -27.7   & -21.1        & -19.0                & -21.9   & 0.4             \\
$7s_{1/2}$ & -42.2                              & -12.2   & -9.5         & -8.3                 & -9.9     & -23.3           \\
$6d_{3/2}$ & -41.5                              & 2.8     & 7.1          & 8.1                  & 6.4     & 6.4             \\
$6d_{5/2}$ & -31.7                              & 12.7    & 18.2         & 19.2                 & 17.1    & 5.2             \\  
\end{tabular}\end{ruledtabular}
\end{table*}

Similarly, Table \ref{tab:lanthanum-energies} shows our calculated energy levels for La$^{2+}$, with comparison to experiment. We can see that there is very close agreement for all the lowest-lying states, with a deviation from experiment at the 0.1\% level or below, except for the $5d$ and $4f$ states. The deviation for the $5d$ states is due to the omission of the ladder diagrams, as discussed for francium. 
The deviation for the $5f_{5/2}$ level from experiment is 835\,cm$^{-1}$, which is of the same order as the Breit correction at the $\Sigma^{(2)}$ level of 748\,cm$^{-1}$. This same behaviour occurs also for ions such as Ce$^{3+}$, Ac$^{2+}$ and Th$^{3+}$, where the $f$ states undergo very large Breit corrections at the $\Sigma^{(2)}$ level, which are of the same order of magnitude as the large deviation from experiment for these states. This is a particularly important problem for systems like Ce$^{3+}$ and Th$^{3+}$, since for these ions the ground states are $4f_{5/2}$ and $5f_{5/2}$, respectively. It can further be seen from Table \ref{tab:lanthanum-energies} that while including Breit into $\Sigma^{(\infty)}$ leads to an important correction, bringing the very large $\Sigma^{(2)}$ Breit correction down from 748.1\,cm$^{-1}$ to 719\,cm$^{-1}$, this is not enough to resolve the large disagreement with experiment. Again, this same behaviour also occurs for Ce$^{3+}$, Ac$^{2+}$ and Th$^{3+}$. Thus, although the very large Breit corrections at second order are somewhat lessened with its inclusion in the all-orders correlation potential, the additional correction is not large enough to resolve the deviation from experiment.
Similarly, large Breit corrections to $f$ states were found in Ref.~\cite{Safronova:2007.76.042504} using a simpler treatment of the Breit interaction. 

One important additional note is that in radium, the $7d_{5/2}$ level has a smaller ionization energy than the $5f_{5/2}$ level but this ordering is incorrectly predicted at the $\Sigma^{(\infty)}$ level; this is corrected with the inclusion of Breit. This is another indication that the Breit interaction plays an important role when included into correlation effects, as this ordering is only corrected only when the Breit interaction is included into $\Sigma^{(2)}$ or $\Sigma^{(\infty)}$. Table \ref{tab:lanthanum-breit-qed-corrections} shows a detailed breakdown of the contributions of Breit at each level of perturbation theory. Here we can more clearly see how the Breit corrections are modified upon its inclusion into $\Sigma^{(\infty)}$, but that it is not enough to greatly reduce the deviation from experiment.  Lastly, we present in Table \ref{tab:cumulative-energy-levels} our calculations for the energy levels of a number of ions along the Cs- and Fr-isoelectronic sequences, with comparison to the experimental values.

\begin{table*}
\squeezetable
\caption{Comparison of theoretical fine structure intervals of La$^{2+}$, and the lowest $f$ intervals of Ce$^{3+}$, Ac$^{2+}$ and Th$^{3+}$, to experimental values (cm$^{-1}$), with the breakdown of the contribution at each level of approximation. The sum of each column gives the total theoretical fine structure interval.}
\label{tab:lanthanum-intervals}
\begin{ruledtabular}\begin{tabular}{@{}llD{.}{.}{6.1}D{.}{.}{2.1}D{.}{.}{2.1}D{.}{.}{3.1}D{.}{.}{3.1}D{.}{.}{6.1}D{.}{.}{6.1}D{.}{.}{4.1}@{}}

       & & \multicolumn{1}{c}{$\Sigma^{(\infty)}$} & \multicolumn{1}{c}{$\delta B^{(\mathrm{DF})}$} & \multicolumn{1}{c}{$\delta B(\Sigma^{(2)})$} & \multicolumn{1}{c}{$\delta B(\Sigma^{(\infty)})$} & \multicolumn{1}{c}{QED}  & \multicolumn{1}{c}{Total}   & \multicolumn{1}{c}{Expt.}   & \multicolumn{1}{c}{$\Delta$} \\ \hline
& $6p$ & 3143             & -30.8                & -5.6       & 1.0           & 5.6 & 3113.1 & 3095.9 & 17.2 \\
& $7p$ & 1199             & -12.6                & -1.5       & -0.6            & 1.5 & 1186.5 & 1229.1 & -42.6  \\
La$^{2+}$ & $5d$ & 1656             & -39.5                & -8.0       & 1.8           & 6.2 & 1616.7 & 1603.2 & 13.5 \\
&$6d$ & 436              & -9.9                 & -1.2       & 0.4           & 1.3 & 427.4  & 433.5  & -6.1   \\
&$4f$ & 1618             & -103.4               & -38.1      & 11.1          & 9.6 & 1497.8 & 1500.3 & -2.5   \\
\hline
Ce$^{3+}$ & $4f$ & -2421 & 166.4 & 28.6 & -9.4 & -13.1 & -2248.2 & -2253.0 & 4.8 \\ \hline
Ac$^{2+}$ & $5f$ & -2675 & 44.7 & 37.9 & -10.3 & -18.1 & -2620.9 & -2625.8 & 4.9 \\ \hline
Th$^{3+}$ & $5f$ & -4433 & 110.5 & 30.7 & -9.5 & -24.0 & -4325.3 & -4325.0 & -0.3 \\
\end{tabular}\end{ruledtabular}
\end{table*}

\subsection{Fine structure intervals}
Table \ref{tab:lanthanum-intervals} shows our calculated fine structure intervals at the $\Sigma^{(\infty)}$ level, with the subsequent columns showing the corrections from including Breit at each order of perturbation theory. Because fine structure splitting is a highly relativistic effect, Breit should have a much larger impact on the accuracy of the fine structure intervals compared to correlation corrections. We can see that for the $p$ and $d$ intervals, the inclusion of Breit at each level generally doesn't improve or worsen the agreement with experiment: for the $6p$, $7p$ and $5d$, the correction to the fine structure interval from the Breit contribution at the $\Sigma^{(\infty)}$ level is away from experiment, while for the $6d$ and $5f$ intervals the correction brings the theoretical value closer to experiment. However, for the $4f$ interval, the $B^{(\infty)}$ correction brings the theoretical prediction within 0.2\% of the experimental value. The $\Sigma^{(\infty)}$ correction is similarly important for the $4f$ interval in Ce$^{3+}$ and the $5f$ interval in Th$^{3+}$, which can be seen at the bottom of Table \ref{tab:lanthanum-intervals}. Therefore, while the energy levels of the $f$ states of moderately charged ions are only marginally improved by the inclusion of the Breit interaction into the all-orders correlation potential, the $f$ fine structure intervals are significantly improved. 

\begin{table*}
\caption{Energy levels for selected Cs- and Fr-like ions (cm$^{-1}$). The first column $\Sigma^{(\infty)}$ shows the energy level using the all-orders correlation potential method, while the subsequent two columns show the resulting corrections from including the Breit interaction and the QED radiative potential method, with the Breit column being the Breit correction at the $\Sigma^{(\infty)}$ level. $\Delta$ shows the percentage deviation from experimental values taken from Ref.\ \cite{NIST_ASD}.}
\label{tab:cumulative-energy-levels}
\begin{ruledtabular}\begin{tabular}{@{}lD{.}{.}{5.0}D{.}{.}{3.1}D{.}{.}{3.1}D{.}{.}{5.0}D{.}{.}{5.0}D{.}{.}{2.2}l|cD{.}{.}{5.0}D{.}{.}{3.1}D{.}{.}{3.1}D{.}{.}{5.0}D{.}{.}{5.0}D{.}{.}{1.2}@{}}
Fr         & \multicolumn{1}{c}{$\Sigma^{(\infty)}$} & \multicolumn{1}{c}{Breit} & \multicolumn{1}{c}{QED}   & \multicolumn{1}{c}{Total} & \multicolumn{1}{c}{Expt.} & \multicolumn{1}{c}{$\Delta$ (\%)} & \hspace{0.3em} & Ra$^+$     & \multicolumn{1}{c}{$\Sigma^{(\infty)}$} & \multicolumn{1}{c}{Breit} & \multicolumn{1}{c}{QED}   & \multicolumn{1}{c}{Total} & \multicolumn{1}{c}{Expt.} & \multicolumn{1}{c}{$\Delta$ (\%)} \\ \hline
$7s_{1/2}$ & 32843               & 0.3   & -45.3 & 32798 & 32849 & -0.15      &   & $7s_{1/2}$ & 81958               & -16.3 & -86.3 & 81855 & 81843 & 0.02          \\
$7p_{1/2}$ & 20625               & -15.6 & 0.9   & 20610 & 20611 & -0.01      &   & $6d_{3/2}$ & 70124               & 66.5  & 42.7  & 70233 & 69758 & 0.68          \\
$7p_{3/2}$ & 18910               & -2.1  & -0.4  & 18908 & 18925 & -0.09      &   & $6d_{5/2}$ & 68438               & 91.5  & 33.1  & 68562 & 68100 & 0.68          \\
$6d_{3/2}$ & 16622               & 29.9  & 11.5  & 16663 & 16619 & 0.27       &   & $7p_{1/2}$ & 60606               & -53.5 & 2.3   & 60555 & 60491 & 0.11          \\
$6d_{5/2}$ & 16440               & 32.9  & 8.9   & 16482 & 16419 & 0.38       &   & $7p_{3/2}$ & 55669               & -13.3 & -1.7  & 55654 & 55634 & 0.04          \\
$8s_{1/2}$ & 13095               & -0.9  & -9.8  & 13085 & 13109 & -0.18      &   & $8s_{1/2}$ & 38427               & -7.1  & -24.7 & 38395 & 38437 & -0.11         \\
$8p_{1/2}$ & 9735                & -5.2  & 0.3   & 9730  & 9736  & -0.06      &   & $7d_{3/2}$ & 33070               & 9.1   & 8.3   & 33088 & 33098 & -0.03         \\
$8p_{3/2}$ & 9187                & -0.9  & -0.1  & 9186  & 9191  & -0.05      &   & $5f_{5/2}$ & 32489               & 127.9 & 30.4  & 32647 & 32855 & -0.63         \\
$7d_{3/2}$ & 8611                & 7.8   & 3.4   & 8622  & 8605  & 0.20       &   & $7d_{5/2}$ & 32577               & 16.2  & 6.5   & 32599 & 32602 & -0.01         \\
$7d_{5/2}$ & 8525                & 9.1   & 2.7   & 8537  & 8516  & 0.25       &   & $5f_{7/2}$ & 32262               & 124.8 & 25.0  & 32411 & 32570 & -0.49         \\
$5f_{7/2}$ & 6951                & 0.4   & 0.0   & 6952  &      &            &  & $8p_{1/2}$ & 31244               & -19.9 & 0.8   & 31225 & 31236 & -0.04         \\
$5f_{5/2}$ & 6949                & 0.9   & 0.1   & 6950  &      &            &  & $8p_{3/2}$ & 29449               & -5.7  & -0.6  & 29443 & 29450 & -0.03         \\
\hline
La$^{2+}$         & \multicolumn{1}{c}{$\Sigma^{(\infty)}$} & \multicolumn{1}{c}{Breit} & \multicolumn{1}{c}{QED}   & \multicolumn{1}{c}{Total} & \multicolumn{1}{c}{Expt.} & \multicolumn{1}{c}{$\Delta$ (\%)} & & Ac$^{2+}$     & \multicolumn{1}{c}{$\Sigma^{(\infty)}$} & \multicolumn{1}{c}{Breit} & \multicolumn{1}{c}{QED}   & \multicolumn{1}{c}{Total} & \multicolumn{1}{c}{Expt.} & \multicolumn{1}{c}{$\Delta$ (\%)} \\ \hline
$5d_{3/2}$ & 155391              & 60.6  & 35.6  & 155487 & 154675 & 0.52     &     & $7s_{1/2}$ & 141159              & -41.6  & -130.0 & 140987 & 140630 & 0.25          \\
$5d_{5/2}$ & 153734              & 106.3 & 29.4  & 153870 & 153072 & 0.52      &    & $6d_{3/2}$ & 140622              & 68.7   & 67.1   & 140758 & 139829 & 0.66          \\
$4f_{5/2}$ & 147476              & 719.1 & 119.1 & 148315 & 147480 & 0.57      &    & $6d_{5/2}$ & 137166              & 120.0  & 51.5   & 137337 & 136426 & 0.67          \\
$4f_{7/2}$ & 145858              & 849.4 & 109.5 & 146817 & 145980 & 0.57      &    & $5f_{5/2}$ & 117194              & 594.4  & 157.6  & 117946 & 117176 & 0.66          \\
$6s_{1/2}$ & 141210              & -22.5 & -68.1 & 141119 & 141084 & 0.02      &    & $5f_{7/2}$ & 114519              & 666.7  & 139.5  & 115325 & 114550 & 0.68          \\
$6p_{1/2}$ & 112788              & -57.3 & 6.0   & 112737 & 112660 & 0.07      &    & $7p_{1/2}$ & 111680              & -103.5 & 3.0    & 111579 & 111164 & 0.37          \\
$6p_{3/2}$ & 109645              & -21.9 & 0.4   & 109624 & 109564 & 0.05      &    & $7p_{3/2}$ & 102949              & -31.8  & -4.0   & 102913 & 102567 & 0.34          \\
$7s_{1/2}$ & 72309               & -9.9  & -23.3 & 72276  & 72328  & -0.07     &    & $8s_{1/2}$ & 71860               & -17.7  & -43.0  & 71800  &        &               \\
$6d_{3/2}$ & 72179               & 6.4   & 6.4   & 72192  & 72294  & -0.14     &    & $7d_{3/2}$ & 68031               & 6.7    & 13.9   & 68051  &        &               \\
$6d_{5/2}$ & 71742               & 17.1  & 5.2   & 71765  & 71861  & -0.13     &    & $7d_{5/2}$ & 66975               & 21.6   & 10.7   & 67007  &        &               \\
\hline
Ce$^{3+}$         & \multicolumn{1}{c}{$\Sigma^{(\infty)}$} & \multicolumn{1}{c}{Breit} & \multicolumn{1}{c}{QED}   & \multicolumn{1}{c}{Total} & \multicolumn{1}{c}{Expt.} & \multicolumn{1}{c}{$\Delta$ (\%)} & & Th$^{3+}$     & \multicolumn{1}{c}{$\Sigma^{(\infty)}$} & \multicolumn{1}{c}{Breit} & \multicolumn{1}{c}{QED}   & \multicolumn{1}{c}{Total} & \multicolumn{1}{c}{Expt.} & \multicolumn{1}{c}{$\Delta$ (\%)} \\ \hline
$4f_{5/2}$ & 298087              & 869.7  & 153.3 & 299110 & 297670 & 0.48      &    & $5f_{5/2}$ & 231687              & 760.3  & 215.1  & 232663 & 231065 & 0.69          \\
$4f_{7/2}$ & 295666              & 1055.2 & 140.2 & 296862 & 295417 & 0.49      &    & $5f_{7/2}$ & 227254              & 892.1  & 191.1  & 228337 & 226740 & 0.70          \\
$5d_{3/2}$ & 248771              & 54.5   & 48.2  & 248874 & 247933 & 0.38      &    & $6d_{3/2}$ & 223190              & 55.8   & 88.9   & 223334 & 221872 & 0.66          \\
$5d_{5/2}$ & 246208              & 122.4  & 39.2  & 246369 & 245444 & 0.38      &    & $6d_{5/2}$ & 217816              & 135.3  & 67.6   & 218018 & 216579 & 0.66          \\
$6s_{1/2}$ & 211329              & -41.8  & -95.4 & 211192 & 211068 & 0.06      &    & $7s_{1/2}$ & 209030              & -73.9  & -177.1 & 208779 & 207934 & 0.41          \\
$6p_{1/2}$ & 174690              & -111.0 & 6.7   & 174586 & 175085 & -0.29     &    & $7p_{1/2}$ & 171877              & -163.4 & 2.8    & 171717 & 170826 & 0.52          \\
$6p_{3/2}$ & 169840              & -48.2  & -0.9  & 169791 & 170378 & -0.34     &    & $7p_{3/2}$ & 158860              & -55.7  & -7.3   & 158797 & 158009 & 0.50          \\
$6d_{3/2}$ & 117790              & -1.2   & 9.9   & 117799 & 120472 & -2.22     &    & $8s_{1/2}$ & 111292              & -34.3  & -62.6  & 111195 & 111443 & -0.22         \\
$6d_{5/2}$ & 117073              & 16.9   & 7.5   & 117097 & 118757 & -1.40     &    & $7d_{3/2}$ & 110011              & -4.9   & 17.9   & 110024 & 111380 & -1.22         \\
$7s_{1/2}$ & 113015              & -19.7  & -34.3 & 112962 & 114168 & -1.06     &    & $7d_{5/2}$ & 108328              & 18.3   & 13.1   & 108359 & 109638 & -1.17         \\
\end{tabular}\end{ruledtabular}
\end{table*}

\begin{table*}[t!]
\caption{Comparison of cumulative first-order and DF-level corrections to valence energy levels of selected ions (cm$^{-1}$) due to the frequency-independent Breit interaction $B$ and the frequency-dependent Breit interaction $B_\omega$.}
\label{tab:dynamic-breit-corrections}
\begin{ruledtabular}\begin{tabular}{@{}lD{.}{.}{3.3}D{.}{.}{3.3}D{.}{.}{3.3}D{.}{.}{3.3}llD{.}{.}{4.2}D{.}{.}{4.2}D{.}{.}{4.2}D{.}{.}{4.2}@{}}
Cs         & \multicolumn{1}{c}{$\langle B\rangle$} & \multicolumn{1}{c}{$\langle B_\omega\rangle$} & \multicolumn{1}{c}{$B^{(\rm{DF})}$} & \multicolumn{1}{c}{$B_\omega^{(\rm{DF})}$}   & \hspace{2em} & Ce$^{3+}$   & \multicolumn{1}{c}{$\langle B\rangle$} & \multicolumn{1}{c}{$\langle B_\omega\rangle$} & \multicolumn{1}{c}{$B^{(\rm{DF})}$} & \multicolumn{1}{c}{$B_\omega^{(\rm{DF})}$} \\ \hline
$6s_{1/2}$ & 29.19            & 29.37                    & 3.20               & 3.88                         &  & $6s_{1/2}$ & 167.9            & 168.4                    & 51.5               & 54.5                       \\
$6p_{1/2}$ & 15.01            & 14.99                    & 7.49               & 7.66                         &  & $6p_{1/2}$ & 167.5            & 167.2                    & 96.3               & 97.6                       \\
$6p_{3/2}$ & 10.84            & 10.49                    & 2.87               & 2.59                         &  & $6p_{3/2}$ & 119.6            & 115.4                    & 47.9               & 44.3                       \\
$5d_{3/2}$ & 12.47            & 12.41                    & -10.15             & -9.85                        &  & $5d_{3/2}$ & 291.9            & 291.0                    & -39.3              & -35.4                      \\
$5d_{5/2}$ & 9.42             & 9.39                     & -11.69             & -11.21                       &  & $5d_{5/2}$ & 219.5            & 219.4                    & -101.0             & -96.1                      \\
$4f_{5/2}$ & 0.00             & 0.00                     & -0.02              & -0.02                        &  & $4f_{5/2}$ & 616.7            & 613.8                    & -806.3             & -789.4                     \\
$4f_{7/2}$ & 0.00             & 0.00                     & -0.02              & -0.02                        &  & $4f_{7/2}$ & 423.6            & 423.2                    & -972.7             & -953.3                     \\ \hline
Fr         & \multicolumn{1}{c}{$\langle B\rangle$} & \multicolumn{1}{c}{$\langle B_\omega\rangle$} & \multicolumn{1}{c}{$\langle B\rangle$} & \multicolumn{1}{c}{$B_\omega^{(\rm{DF})}$} &  & Th$^{3+}$   & \multicolumn{1}{c}{$\langle B\rangle$} & \multicolumn{1}{c}{$\langle B_\omega\rangle$} & \multicolumn{1}{c}{$B^{(\rm{DF})}$} & \multicolumn{1}{c}{$B_\omega^{(\rm{DF})}$} \\ \hline
$7s_{1/2}$ & 66.94            & 67.39                    & 6.34               & 10.15                        &  & $7s_{1/2}$ & 324.8            & 327.3                    & 88.5               & 102.3                      \\
$7p_{1/2}$ & 29.63            & 29.80                    & 14.26              & 15.39                        &  & $7p_{1/2}$ & 298.5            & 301.3                    & 164.3              & 173.6                      \\
$7p_{3/2}$ & 18.50            & 17.16                    & 4.40               & 3.26                         &  & $7p_{3/2}$ & 184.1            & 169.9                    & 66.6               & 54.3                       \\
$6d_{3/2}$ & 18.81            & 18.72                    & -12.50             & -11.46                       &  & $6d_{3/2}$ & 438.9            & 436.3                    & -42.9              & -30.1                      \\
$6d_{5/2}$ & 14.90            & 14.76                    & -14.08             & -12.45                       &  & $6d_{5/2}$ & 324.8            & 322.3                    & -111.3             & -97.8                      \\
$5f_{5/2}$ & 0.01             & 0.01                     & -0.05              & -0.05                        &  & $5f_{5/2}$ & 704.5            & 702.1                    & -698.1             & -660.2                     \\
$5f_{7/2}$ & 0.01             & 0.01                     & -0.05              & -0.04                        &  & $5f_{7/2}$ & 520.2            & 520.3                    & -808.6             & -770.0                     \\ 
\end{tabular}\end{ruledtabular}
\end{table*}

\section{Additional relativistic corrections}
It is clear that while including the (one-body) Breit interaction into the all-orders correlation potential improves fine structure intervals, it does not meaningfully contribute to resolving the large deviation from experiment between the theoretical and experimental energies of Cs- and Fr-like ions. Given that the Breit corrections seem to be the right order of magnitude to resolve this issue, we can therefore try including additional relativistic corrections beyond the Breit interaction. In theory this could be done by treating the exchange of a single photon to second order in QED and retaining terms of order $(v/c)^3$. However, it is possible to retain all terms in the relativistic expansion of the photon propagator, and this leads to a generalization of Eq.\ (\ref{eq:breit-interaction}) known as the frequency-dependent Breit interaction. When the photon field is quantized in the Coulomb gauge, the resulting expression is
\begin{align}
\begin{split}
    B_C(\bm{r}_1,\bm{r}_2;&\,\omega)=-e^2\Bigg\{\frac{\bm{\alpha}_1\cdot\bm{\alpha}_2}{r}  \\
    &+(\bm{\alpha}_1\cdot\nabla_1)(\bm{\alpha}_2\cdot\nabla_2)\left[\frac{\cos(\omega r) - 1}{\omega^2r}\right]\bigg\},
\end{split}
\end{align}
while in the Lorenz gauge the analogous result is \cite{Mann:1971.4.41, Grant:2007kae}
\begin{align}
    B_L(\bm{r}_1,\bm{r}_2;\omega)&=-\frac{e^2}{r}\left[(\bm{\alpha}_1\cdot\bm{\alpha}_2 - 1)\cos(\omega r) + 1\right],
\end{align}
which is also sometimes called the M\o ller interaction. It can be shown that matrix elements of the two different operators in the different gauges agree only when the wave functions are solved in a local potential, and therefore are not equal when the DF approximation is used in calculating the wave functions. However, it is generally accepted that the Coulomb gauge expression is the appropriate one to use, as the Lorenz gauge expression leads to erroneous terms that are only corrected at higher orders in perturbation theory \cite{Chantler:2014.90.062504,Lindgren:1990.23.1085}, and it will be the one that we use here. While several numerical studies have been performed using the frequency-dependent Breit interaction as a first-order perturbation, there are few studies which have included the frequency-dependent Breit interaction into the Dirac-Fock procedure. The contribution to the DF energies from using $B_C(\omega)$ rather than $B$ have been predicted to be small in Ref.~\cite{Lindroth:1989.22.2447}.

\subsection{Numerical results}
Table \ref{tab:dynamic-breit-corrections} shows the effect that including the frequency-dependent Breit interaction  into the Dirac-Fock procedure makes to the energy levels of heavy ions. In this table we also show the correction to the same energy levels by treating $B_C(\omega)$ simply as a first-order perturbation to the system. Firstly, we can see that including the Breit interaction into the DF procedure compared to treating it as a perturbation leads to very different corrections, both in terms of order of magnitude and, for the $d$ and $f$ states, also in terms of sign. For example, in Fr, the corrections to the $5d$ levels that come from treating $B$ in first-order perturbation theory versus in the DF procedure are equal in magnitude but opposite in sign. In some cases this difference is even larger; in Ce$^{3+}$ the DF corrections to the $4f$ levels (both for $B$ and $B_{\omega}$) are opposite in sign but twice as large as the first order corrections.

In comparing the first-order corrections due to $B$ and $B(\omega)$, we can see that the difference is generally very small, being on the order of $\sim0.01\%$ of the $\langle B\rangle$ correction, although this does get larger as we go to heavier ions, which has been observed in Refs.~\cite{Mann:1971.4.41,Grant:1980.13.2671} for the ground state energy of closed-shell atoms. This pattern also occurs for the corresponding DF corrections, although compared to the first order corrections the difference is more noticeable. For example, for the $5f_{5/2}$ level in Th$^{3+}$, the difference between $\langle B_\omega\rangle$ and  $\langle B\rangle$ is just 0.34\% of the total $B_{\omega}$ Breit correction (i.e.\ including frequency-dependence) at this order, while at the DF level the difference is 5.7\% of the total $B_{\omega}^{(\rm{DF})}$ correction. Therefore, the frequency-dependence is more important when Breit is included into the DF procedure than it is when it is just treated as a first order perturbation.  

Despite this, what is important for our purposes is that the difference between including $B$ versus $B(\omega)$ into the DF procedure is very small for the Cs-like ions, and while it becomes larger for the Fr-like ions, it does not seem to be of the appropriate size to resolve the large discrepancy that $B$ seems to introduce into the calculation of energy levels. For example, the DF correction to the $5f_{5/2}$ level due to $B$ is $-698.1$\,cm$^{-1}$, while the corresponding correction due to $B(\omega)$ is somewhat smaller at $-660.2\,$cm$^{-1}$. While this correction does bring the theoretical value closer to the experimental value, it is still not of the right size to mend the large discrepancy between them.

\section{Conclusion}
In this work we have demonstrated the effect that the Breit interaction plays in the calculation of atomic properties of moderately-charged, heavy ions when included into the all-orders correlation potential method. In the Fr- and Cs-isoelectronic sequences, the effect of the Breit interaction at second order in the residual electron interaction to the energy levels of $f$ states is large, and is of the same order of magnitude as the significant deviation from experiment between the theoretical and experimental values of these energies. Upon including the Breit interaction into the Feynman Green's function in the all-orders correlation potential method, we find excellent agreement between theory and experiment for the fine structure intervals; however, we find that the large deviation from experiment for the energy levels remains. Furthermore, we include the frequency-dependent Breit interaction into the Dirac-Fock equations, and we find that this is a negligible contribution, as expected.

\acknowledgments
\noindent
This work was supported by Australian Research Council through DECRA Fellowship DE210101026, and Discovery Project DP230101685.

\bibliography{references, references2}

\end{document}